# Phonon Transport within Periodic Porous Structures — From Classical Phonon Size Effects to Wave Effects

Yue Xiao[1], Qiyu Chen[1], Dengke Ma[2,3], Nuo Yang[2,4*], Qing Hao[1,*]

1. Department of Aerospace and Mechanical Engineering, University of Arizona, Tucson, AZ 85721 U.S.A
2. State Key Laboratory of Coal Combustion, Huazhong University of Science and Technology (HUST), Wuhan 430074, P. R. China
3. NNU-SULI Thermal Energy Research Center (NSTER) & Center for Quantum Transport and Thermal Energy Science (CQTES), School of Physics and Technology, Nanjing Normal University, Nanjing 210023, P. R. China
4. Nano Interface Center for Energy(NICE), School of Energy and Power Engineering, Huazhong University of Science and Technology (HUST), Wuhan 430074, P. R. China

[*]Corresponding authors: Qing Hao (qinghao@email.arizona.edu), Nuo Yang (nuo@hust.edu.cn)

## Nomenclature

| | |
|---|---|
| $A$ | Pore surface area. For a cylindrical pore, this is the sidewall surface area of its inner wall. |
| $C$ | Phonon specific heat |
| $d$ | Pore diameter |
| $F(\Phi)$ | Fourier-law-based correction factor to account for the thermal conductivity reduction due to the porosity $\Phi$ |
| $f$ | Phonon frequency |
| $I_{b\eta}$ | Intensity for wavenumber $\eta$ |
| $k$ | Thermal conductivity |
| $k_E$ | Electronic thermal conductivity |
| $k_L$ | Lattice thermal conductivity |
| $L$ | Characteristic length of nanoporous films, which is determined by the geometry of the film. |
| $L_{eff}$ | Effective characteristic length of nanoporous films, which is determined by matching the lattice thermal conductivity predicted by the phonon Monte Carlo simulations and that by the kinetic relationship using $L_{eff}$ to modify bulk phonon mean free paths. |
| $L_0$ | Metallic Lorenz number |
| $L_d(\hat{s})$ | Distance along a line to a point at the opposing surface along a unit direction vector $\hat{s}$ within an enclosure |
| $p$ | Pitch, as the center-to-center distance between adjacent periodic pores |
| $P$ | Perimeter |
| $P(\lambda)$ | Specularity of film-surface phonon reflection for a given phonon wavelength $\lambda$ |
| $q_\eta$ | Heat flux for wavenumber $\eta$ |
| $S$ | Area of 2D structures |
| $S_{Solid}$ | Solid area of a period |
| $\hat{s}$ | Unit vector for traveling direction |
| $T$ | Absolute temperature |
| $t$ | Film thickness |
| $V$ | Volume |
| $V_{Solid}$ | Solid-region volume in a nanoporous structure |
| $v_{avg}$ | Averaged sound velocity |
| $v_g$ | Phonon group velocity |

## Greek Symbols

| | |
|---|---|
| $\delta$ | Boundary/interface roughness |
| $\eta$ | Wave number |
| $\eta_{wave}$ | Wave ratio, the fraction of thermal conductivity reduction due to wave effects. |
| $\theta$ | Angle related to the phonon traveling direction |
| $\kappa_\eta$ | Spectral absorption coefficient |
| $\Lambda_{bulk}$ | Bulk phonon mean free path |
| $\Lambda_{eff}$ | Effective phonon mean free path |
| $\Lambda_{Film}$ | Effective phonon mean free path for the in-plane heat conduction along a thin film |
| $\Lambda_{Pore}{}^*$ | Dimensionless characteristic length of a nanoporus material |
| $\lambda$ | Phonon wavelength |
| $\sigma$ | Electrical conductivity |
| $\Phi$ | Porosity, as the volumetric percentage of nanopores |
| $\Omega$ | Solid angle |

## Subscripts

| | |
|---|---|
| GPnC | Graphene phononic crystal |
| SiNW | Silicon nanowire |
| NCJ_MC | Nano-cross-junction system with its thermal conductivity calculated by the Monte Carlo method |
| NCJ_AGFMC | Nano-cross-junction system with its thermal conductivity calculated by the AGFMC method |

## Abstract

Tailoring thermal properties with nanostructured materials can be of vital importance for many applications. Generally classical phonon size effects are employed to reduce the thermal conductivity, where strong phonon scattering by nanostructured interfaces or boundaries can dramatically supress the heat conduction. When these boundaries or interfaces are arranged in a periodic pattern, coherent phonons may have interference and modify the phonon dispersion, leading to dramatically reduced thermal conductivity. Such coherent phonon transport has been widely studied for superlattice films and recently emphasized for periodic nanoporous patterns. Although the wave effects have been proposed for reducing the thermal conductivity, more recent experimental evidence shows that such effects can only be critical at an ultralow temperature, i.e., around 10 K or below. At room temperature, the impacted phonons are mostly restricted to hypersonic modes that contribute little to the thermal conductivity. In this review, the theoretical and experimental studies of periodic porous structures are summarized and compared. The general applications of periodic nanostructured materials are further discussed.

## 1. Introduction — Challenges in Manipulating Heat as Waves

Nanostructured materials introduce unique opportunities to tailor the intrinsic transport properties. In structures such as solid thin films, the reduced lattice thermal conductivity ($k_L$) is often attributed to the boundary scattering of particle-like phonons, as the classical phonon size effect. When periodic nano-patterns are introduced within a structure, unique opportunities also exist in using the wave nature of lattice vibrations to modify the phonon dispersion and thus the $k_L$. The overall thermal conductivity $k$, consisting of both $k_L$ and the electronic contribution $k_E$, can also be largely suppressed. This can benefit applications requiring a low $k$, such as thermoelectric energy conversion and thermal insulation materials.[1] As one important research direction of reducing the thermal transport, periodic nanoporous thin films[2-18] and graphene[19, 20] (also known as graphene antidot lattices or GALs[21], as shown in Fig. 9a) have been widely studied.

In above mentioned 2D periodic porous structures, possibly coherent interference between lattice vibration waves can lead to opened phononic bandgaps. The modified phonon density of states (DOS) and phonon group velocities can largely reduce $k_L$. In analogy to photonic crystals with periodic cavities to manipulate light,[22, 23] periodic nanoporous films and antidot lattices are referred to as "phononic crystals." Such phononic effects has been intensively studied for superlattice thin films with atomically smooth interfaces between alternating layers, in which coherent phonon transport becomes dominant for <5 nm periods at 300 K.[24, 25] In comparison, nanopores within thin films or GALs may play the same role as interfaces within superlattices. Their periodicity modifies phonon dispersions, whereas their rough edges act as defects to scatter phonons. However, it is more challenging to observe room-temperature phononic effects in nanoporous films due to the technical difficulty in achieving: 1) ultrafine patterns with sub-10 nm periodic length or pitch $p$; and 2) smooth pore edges with minimized defects. As the guidance for phononic studies, these two requirements are discussed in details below.

For the first requirement, the sub-10 nm pitch is necessary to match the short wavelength of majority phonons at room temperature (1–10 nm at 300 K for bulk Si[26, 27]) because the impacted phonons should have wavelengths comparable to or shorter than the periodic length of the structure. Similar to the Bragg gaps for photonic crystals, the frequency of phonons impacted by the periodic structure can be approximated as $f = \pi v_{avg}/p$, in which $v_{avg}$ is the averaged sound velocity.[6] For $p$=100 nm, phonons with $f$~200 GHz or below can be impacted, instead of majority heat-carrying phonons at a few THz in bulk Si. The overall impact on $k_L$ is very limited. In the literature, the smallest $p$ is 34 nm for measured nanoporous Si films.[14] Further reducing this $p$ value to 10 nm or below is limited by the ~5 nm spatial resolution of the state-of-the-art electron beam lithography (EBL). For a film with its thickness $t$, the smallest pore diameter $d$ is also restricted by the aspect ratio $t/d < 3$ for dry etching, i.e., $d > t/3$.[10] For atomic-thick materials, the limitation due to the aspect ratio is removed but accurate thermal measurements still remain as a challenge.

For the second requirement, pore edges should be smooth enough to keep the phase information of coherent phonons before and after the pore-edge scattering. Phononic effects require specular phonon reflection on all boundaries because diffuse phonon scattering will destroy the coherent phonon phase.[28] In Ziman's theory, the probability $P(\lambda)$ for specular reflection can be estimated as $P(\lambda) = exp[-16\pi^2(\delta/\lambda)^2]$, in which $\delta$ is the average boundary roughness and $\lambda$ is phonon wavelength.[29,30] Following this $P(\lambda)$ expression, even $\delta$ as small as 1 nm can yield completely diffusive phonon scattering to destroy phononic effects. In real samples, a layer of 1–2 nm native oxide on the pore sidewalls is revealed by high-resolution transmission electron microscopy (TEM).[31] Such amorphous pore edges are inevitable due to the structure damage by the pore-drill process and strong oxidation of nanostructured surfaces. Further considering the diffusive phonon refection on the top and bottom surfaces of a thin film, strong phonon coherence at 300 K is not anticipated for reported periodic nanoporous Si thin films. In this aspect, the claimed room-temperature phononic effects for pores with 2.5 nm surface roughness[2] may be attributed to measurement errors using a microdevice.

In principle, coherent phonon transport also requires structure sizes to be much smaller than majority phonon mean free paths (MFPs) because the internal phonon scattering inside material can also destroy the coherent phonon phase.[32] This issue is less critical because significant percentage of heat is still carried by phonons with very long MFPs. At 300 K, first-principles calculations[26] suggest that 50% of the room-temperature $k_L$ is contributed by phonons with MFPs longer than 1 µm. Along this line, the phonon MFP comparison with $p$ has also been proposed to justify the importance of phononic effects.[2, 4] However, this argument is inconsistent with the understanding from comparable photonic crystal, where the wavelength comparison with $p$ is always used to justify the wave effect.

Compared with photons, phonons with much shorter wavelengths are more difficult to be manipulated for their wave effects using periodic nanostructures. This article reviews existing studies on atomic to nanoscale periodic porous structures to tune the phonon transport. The limitation and important applications of phononic effects are discussed. The discussions cover two-dimensional (2D) periodic porous thin films and GALs, and three-dimensional (3D) porous nano-cages.

## 2. Periodic nanoporous films (quasi-2D phononic crystal)

For widely studied periodic nanoporous Si films, phononic effects are negligible above room temperature in measured samples because the diffusive pore-edge phonon scattering can destroy the coherent phonon phase and large feature sizes have limited impact on majority heat-carrying phonons. In existing studies, phononic effects can only be confirmed at cryogenic temperatures, where the dominant phonon wavelengths scale up with $1/T$, with $T$ as the absolute temperature.[33] Such trends are consistent with recent measurements with phononic effects determined below 14 K for Si films with $p$>100 nm,[34] or below 10 K for $p$=300 nm.[35] In nanoporous $SiN_x$ films, strong phononic effects have been observed at sub-Kelvin temperatures for a pitch $p$ of 970 and 2425 nm.[18]

In most cases, the classical phonon size effects are still the major mechanism for the thermal conductivity reduction. The $k_L$ predictions are mainly based on the Boltzmann transport equation (BTE) assuming incoherent phonon transport. For ultrafine nanoporous patterns, possible phononic effects can be revealed by molecular dynamics (MD) simulations.[36] The experimental studies are summarized for periodic nanoporous thin films, with a focus on how to justify the phononic effects. Most discussions are based on the in-plane $k_L$ of nanoporous Si thin films. Fewer studies can be found for anisotropic $k_L$ calculations[37, 38] and cross-plane $k_L$ measurements.[7, 39]

### 2.1 Modeling the in-plane $k_L$ of periodic nanoporous thin films

For nanoporous thin films with incoherent phonon transport, accurate in-plane $k_L$ can be given by phonon Monte Carlo (MC) simulations[40-42] that track the transport of individual phonons and statistically obtain the solution of the phonon BTE. Other than complicated phonon MC simulation, analytical modelling has also be carried out using an effective phonon MFP ($\Lambda_{eff}$) in the kinetic relationship:

$$k_L = F(\Phi) C v_g \Lambda_{eff}/3 \, . \tag{1}$$

Here the phonon specific heat $C$ and phonon group velocity $v_g$ are unchanged from the bulk values. The additional factor $F(\Phi)$ accounts for the heat transfer reduction due to the porosity $\Phi$, whereas $\Lambda_{eff}$ further addresses the phonon size effects. In many studies, $k_L/F(\Phi)$ is compared to that of the starting solid thin film to show the phonon size effects due to introduced nanoporosity. In accurate analysis, $k_L$ is integrated over the whole phonon spectrum and summed over different phonon branches. The complicated full phonon dispersion can also be considered.[27]

#### 2.1.1 Correction factor $F(\Phi)$

In principle, $F(\Phi)$ can be determined by the Fourier's law analysis, e.g., as the ratio between the thermal conductances of a porous film and its nonporous counterpart.[27, 42] Analytical expressions of $F(\Phi)$ are also available. In early studies, the Eucken's factor is used and is given as

$$F(\Phi) = \frac{1-\Phi}{1+\Phi/2}, \tag{2}$$

which was derived for a bulk material with cubically aligned spherical pores.[43] For nanoporous films, however, the Hashin-Shtrikman factor[44] is found to be more accurate. This factor is expressed as

$$F(\Phi) = \frac{1-\Phi}{1+\Phi}. \tag{3}$$

In fact, Eq. (3) can also be found as a special case for a 2D composite, where pores corresponds to circular inclusions with $k = 0$.[45] The validation of Eq. (3) with the Fourier's law analysis can be found in some studies.[27, 42]

#### 2.1.2 $\Lambda_{eff}$ calculations using an analytical characteristic length $L$

As another important aspect for the analytical modelling, the effective phonon MFP $\Lambda_{eff}$ can be computed with a MC technique based on ray-tracing or path sampling,[34, 46-48] or solving a MFP-dependent phonon BTE.[49, 50] Considering phonon scattering at pore edges and film surfaces, these techniques should be applied to individual bulk phonon MFP ($\Lambda_{Bulk}$) to find its corresponding $\Lambda_{eff}$. This can be extremely expensive for calculations, considering the wide $\Lambda_{Bulk}$ distribution in materials like Si.[26] To simplify, $\Lambda_{eff}$ is also modified from $\Lambda_{Bulk}$ based on Matthiessen's rule, i.e., $\Lambda_{eff} = (1/\Lambda_{Bulk} + 1/L)^{-1}$.[2, 38, 41, 42, 51] Here a characteristic length $L$ of the nanoporous structure is introduced. A similar treatment can be found for nanowires, where the nanowire diameter is simply the characteristic length to account for the completely diffusive phonon boundary scattering.[52]

For periodic nanoporous films, there are mainly three key geometry parameters: pore pitch $p$ as the averaged center-to-center distance between adjacent pores, film thickness $t$ and pore diameter $d$. In typical patterns, pores are distributed on a square lattice or a hexagonal lattice. For ultrafine nanoporous patterns, the impact of the film thickness can be negligible. Only considering $d$ and $p$, various characteristic lengths have been proposed for periodic nanoporous Si films. Table 1 lists $L$ expressions proposed in different studies. Among these expressions, the neck width $L = p - d$ can be used to predict the lower bound of $k_L$. This $L$ is not expected to be accurate for pores on a square pattern because the second-nearest-neighbor pores have an expanded neck width as $L = \sqrt{2}p - d$ from $L = p - d$ for nearest-neighbor pores.[53] At the limit $p \to d$, however, experimental studies by Anufriev et al.[54] and Yanagisawa et al.[3] suggested that $L = p - d$ became more accurate and the surface roughness was also critical. Calculations by Yu et al.[55] also indicated the importance of the neck width. In the reviews by Marconnet et al.[10] and Nomura et al.,[56] the measured $k$ values in existing studies are plotted as a function of $L = p - d$ for comparison purposes.

At the ballistic limit, it has been found that the geometric mean beam length (MBL) for optically thin media in radiation[41, 42] is simply the accurate characteristic length. The MBL is identical with the results provided by MC ray tracing (MCRT), as introduced by Lacroix et al.[57] to determine the characteristic length of a periodic nanoporous structure. For circular pores, the MBL can be computed as[39, 51]

$$L = \frac{4V_{\text{Solid}}}{A} = \frac{4V(1-\Phi)}{A} = \begin{cases} \frac{4p^2 - \pi d^2}{\pi d} \text{ (square lattice)} \\ \frac{2\sqrt{3}p^2 - \pi d^2}{\pi d} \text{ (hexagonal lattice)} \end{cases}, \quad (4)$$

where the solid-region volume $V_{Solid}$ and pore surface area $A$ are evaluated within a period. For a through-film pore, $A$ is simply $\pi dt$ as the sidewall surface area of a pore. Here $V_{Solid}$ is diverged from the volume $V$ by a factor of $(1 - \Phi)$, with $\Phi$ as the porosity of the structure.

**Table 1.** Characteristic length $L$ of periodic nanoporous Si films. Here $V$ is the volume of one period, including the pore volume. For the pore within this period, $A = (\pi d)t$ is its sidewall surface area.

| Article | $L$ expression | Physical meaning |
|---|---|---|
| Hopkins et al.[58, 59] | $p - d$ | Neck width between adjacent pores |
| Alaie et al.[2] | $\sqrt{p^2 - \pi d^2/4}$ | The square root of the solid area within a period |
| Hao et al.[42] | $4V(1 - \Phi)/A$ | Geometric MBL |
| Huang et al.[38] | $d/8\Phi$ | Traveling distance of a particle to encounter a pore within a swept volume with $td$ cross-section area |

The characteristic length $L$ has also been derived for 2D porous films by Huang et al.[38] and 3D particle-in-a-host composites by Minnich and Chen.[60] In principle, nanoporous materials are viewed as a special case for a particle-in-a-host composite, with zero heat conduction inside the embedded nanoparticles. In a nanocomposite, $A$ becomes the interface area between a particle and the host and $\Phi$ becomes the volumetric percentage of nanoparticles. These two studies[38, 60] both follow the MFP calculations of gas molecules.[61] For the host region of a 3D particle-in-a-host composite or the solid region of a 2D porous film, the derived $L$ is proportional to $V/A$, which is in contrast with $L \sim V(1 - \Phi)/A$ in Eq. (4). This additional $(1 - \Phi)$ factor has been proposed by Machrafi and Lebon[62] to better explain the trend of $L \to 0$ at $\Phi \to 1$. In the literature, the volumetric surface or interface area $A/V$ is acknowledged as one key parameter for $k_L$ reduction.[60, 63] However, the examination here suggests $A/V(1-\Phi)$ or the MBL as a more accurate parameter for $k_L$ modelling. In general, analysis based on the surface-to-volume ratio $A/V$ tends to overpredict $k_L$ but this overprediction can be reduced using the MBL.[41, 42]

### 2.1.3 Effective characteristic length *L*

In practice, none of the characteristic lengths listed in Table 1 can be accurate across the whole phonon MFP spectrum. An effective characteristic length $L_{eff}$ is often used to match $k_L$ predicted by Eq. (1) and $k_L$ yielded by the phonon BTE. Here $\Lambda_{eff} = \left(1/\Lambda_{Bulk} + 1/L_{eff}\right)^{-1}$ is used in Eq. (1) and $L_{eff}$ is fitted. Although the MBL or $L$ given by MCRT is accurate for the ballistic regime, the divergence between $L_{eff}$ and MBL is expected because it is not accurate to use the Matthiessen's rule to combine the boundary phonon scattering on pore edges and internal phonon scattering inside the volume.[61] For circular or square pores on a square lattice, Fig. 1 compares the MBL and effective characteristic length $L_{eff}$ for nanoporous Si films. The phonon BTE solution is given by frequency-dependent phonon MC simulations.[42] Similar $L_{eff}$ values have also been extracted in a separated study.[37]

In Fig. 1, both MBL and $L_{eff}$ are divided by the pitch $p$ to get a dimensionless $\Lambda_{Pore}{}^*$. As a factor only depending on the geometry, $\Lambda_{Pore}{}^* = MBL/p$ is only a function of the porosity. Due to its dependence on the actual phonon MFP distribution relative to the structure size, $\Lambda_{Pore}{}^* = L_{eff}/p$ varies slightly for $p$=50, 200 and 500 nm. The plotted $\Lambda_{Pore}{}^* = L_{eff}/p$ is averaged over these three $p$ values. In Fig. 2, $k_L$ for 2D nanoporous Si thin films with smooth top/bottom surfaces is predicted using the MBL or $L_{eff}$ to modify $\Lambda_{Bulk}$ and thus obtain $\Lambda_{eff}$ in Eq. (1). It can be observed that the MBL typically leads to an overpredicted $k_L$, whereas the extracted $L_{eff}$ can give accurate results for $p$ from 50 to 500 nm.

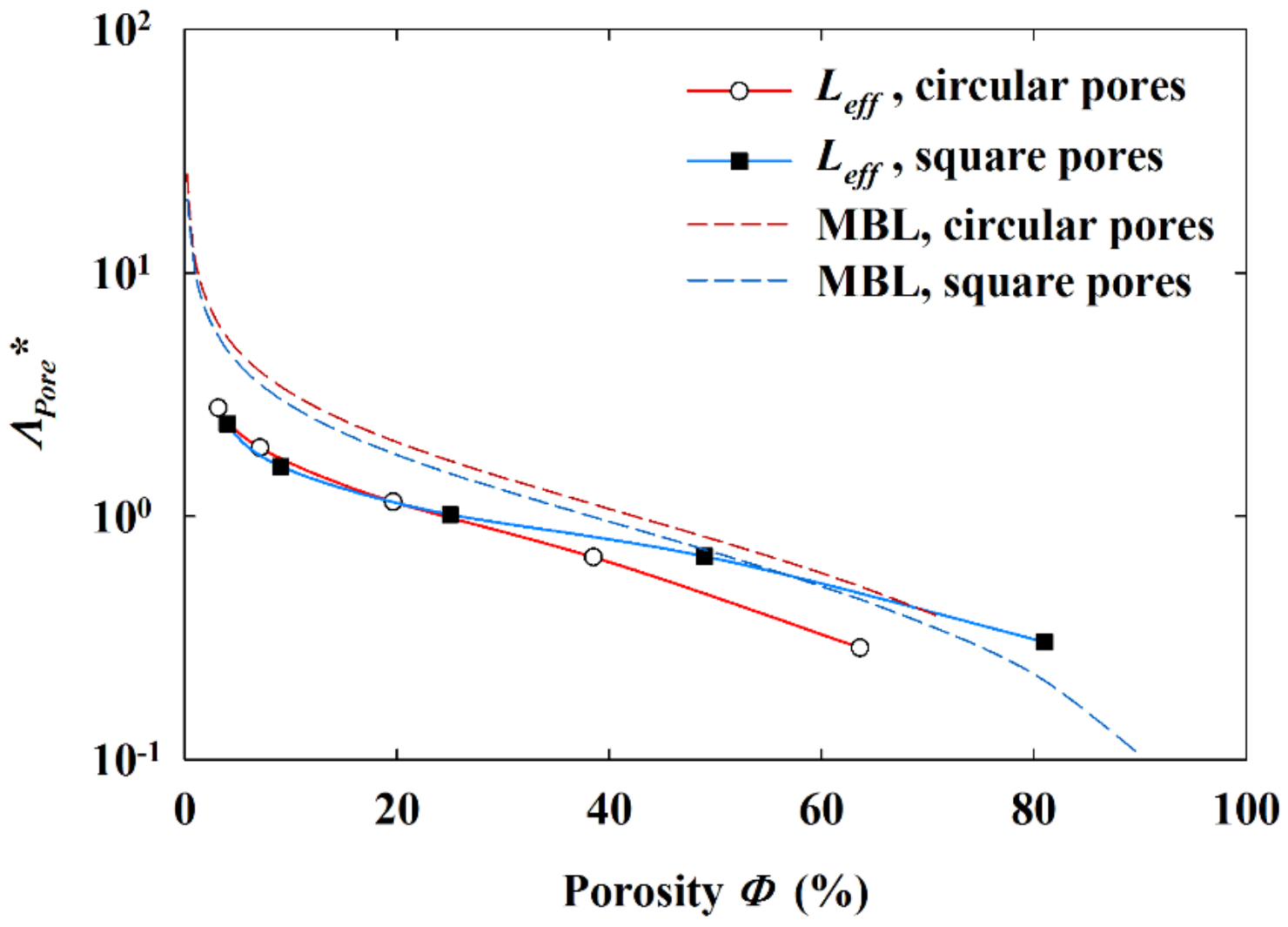

**Fig. 1** Porosity- and period-dependent dimensionless $\Lambda_{Pore}{}^* = L/p$, with $L$ as the MBL or $L_{eff}$. Summarized from figures in Ref. [42]. Copyright 2016 American Institute of Physics.

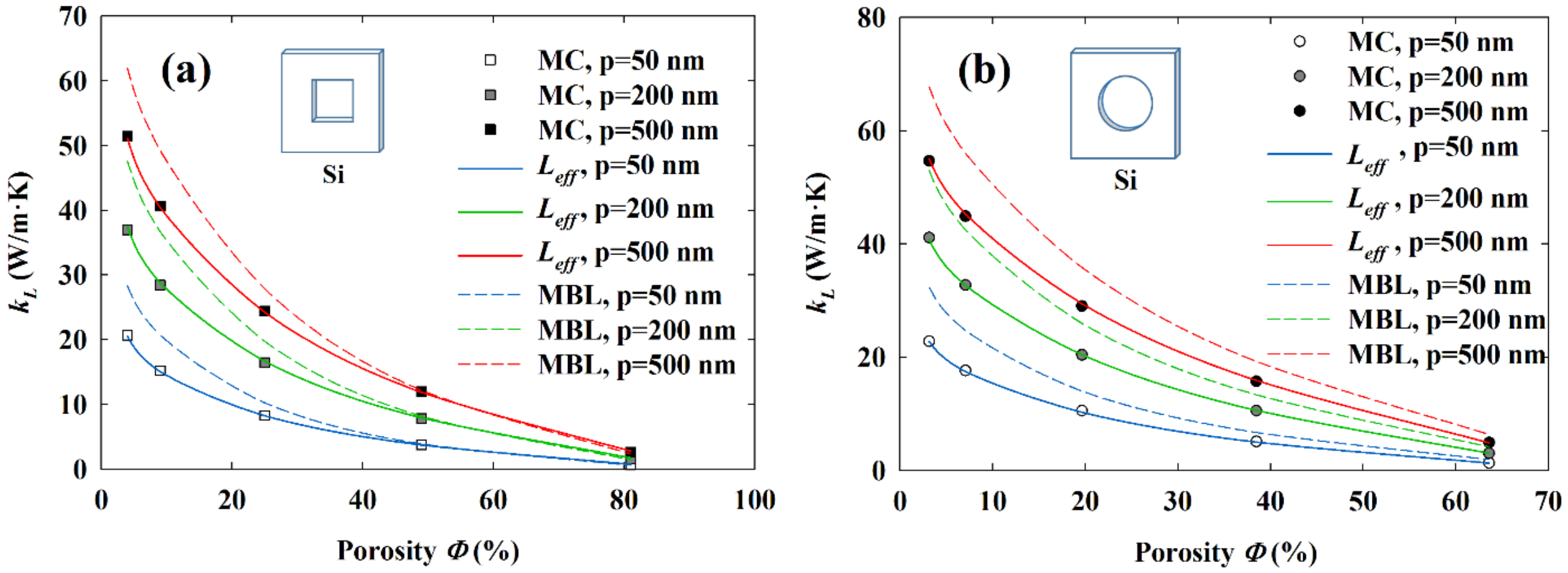

**Fig. 2** Predicted room-temperature $k_L$ of thin films with aligned pores, using $L_{eff}$ (solid lines) and the MBL (dashed lines): (a) Square pores in Si, (b) circular pores in Si. The symbols are predictions from phonon MC simulations to compare with. Reproduced from Ref. [42]. Copyright 2016 American Institute of Physics.

### 2.1.4 Influence of the film thickness

In addition to the characteristic length of a 2D periodic porous pattern, the film thickness $t$ can further influence the in-plane $k_L$ by diffusive film-surface phonon scattering. In one method, Huang et al.[38] modify the bulk phonon MFPs with $t$ and $L$, given as $1/\Lambda_{eff} = 1/\Lambda_{Bulk} + 1/L + 2cos\theta/t$. Here $\theta$ is the included angle between the phonon traveling direction and the cross-plane direction. This $\Lambda_{eff}$ is then used in the solid-angle-dependent integration of $k_L$.

In this review, a two-step phonon MFP modification is proposed to compute $\Lambda_{eff}$, which also offers more flexibility in handling independent interface/boundary scattering processes. First, the bulk phonon MFP $\Lambda_{Bulk}$ can be modified as the in-plane phonon MFP $\Lambda_{Film}$ for a solid film. Based on the Fuchs-Sondheimer model, $\Lambda_{Film}$ is given as[61]

$$\frac{\Lambda_{Film}}{\Lambda_{Bulk}} = 1 - \frac{3[1-P(\lambda)]\Lambda_{Bulk}}{2t}\int_0^1 (x - x^3)\frac{1-\exp\left(-\frac{t}{\Lambda_{Bulk}x}\right)}{1-P(\lambda)\exp\left(-\frac{t}{\Lambda_{Bulk}x}\right)}\mathrm{d}x, \quad (5)$$

where $P(\lambda)$ is the specularity of film-surface phonon reflection. Second, this $\Lambda_{Film}$ can be modified again to obtain an effective phonon MFP for a nanoporous film, i.e., $\Lambda_{eff} = \left(1/\Lambda_{Film} + 1/L_{eff}\right)^{-1}$, using the characteristic length $L_{eff}$ in Fig. 1 for aligned pores. When $L_{eff}$ is unavailable for the nanoporous pattern, the MBL can always be used but some errors are anticipated. Above two steps for the phonon MFP modification address the phonon size effects for the cross-plane direction and in-plane direction, respectively. Finally, $\Lambda_{eff}$ is used in Eq. (1) to compute $k_L$. In frequency-dependent phonon studies, $k_L$ is integrated across the whole phonon spectrum and summed over different phonon branches. Similar two-step phonon MFP modifications from $\Lambda_{Bulk}$ to $\Lambda_{eff}$ can be found in other studies by Hao et al., such as nanoporous thin films with patterned nanoslots[64] and nanograined bulk material with nano-inclusions within each grain.[65] This procedure can be easily extended to general analysis using the exact phonon dispersions and first-principles-computed phonon MFPs.

In demonstration, the analytical model discussed above is used to compute the room-temperature $k_L$ for a 220-nm-thick periodic nanoporous Si film, where MBLs are used as the characteristic length (Fig. 3a). Completely diffusive phonon reflection is assumed on film surfaces, which is a reasonable assumption above 300 K. In comparison, $k_L$ is also predicted by phonon MC simulations considering periodic circular through-film pores[66] (symbols). All calculated $k_L$ are normalized by the Hashin-Shtrikman factor[44] in Eq. (3) to remove the influence of the porosity, yielding $k_L$ for the solid or non-porous counterpart of the thin film. Energy-dependent phonon MFPs fitted for bulk Si[67] are employed here. It can be observed that the MBL as the characteristic length can slightly overpredict the $k_L$. Using the $L_{eff}$ as the characteristic length instead, the predicted $k_L$ becomes consistent with those given by the phonon MC simulations (Fig. 3b).

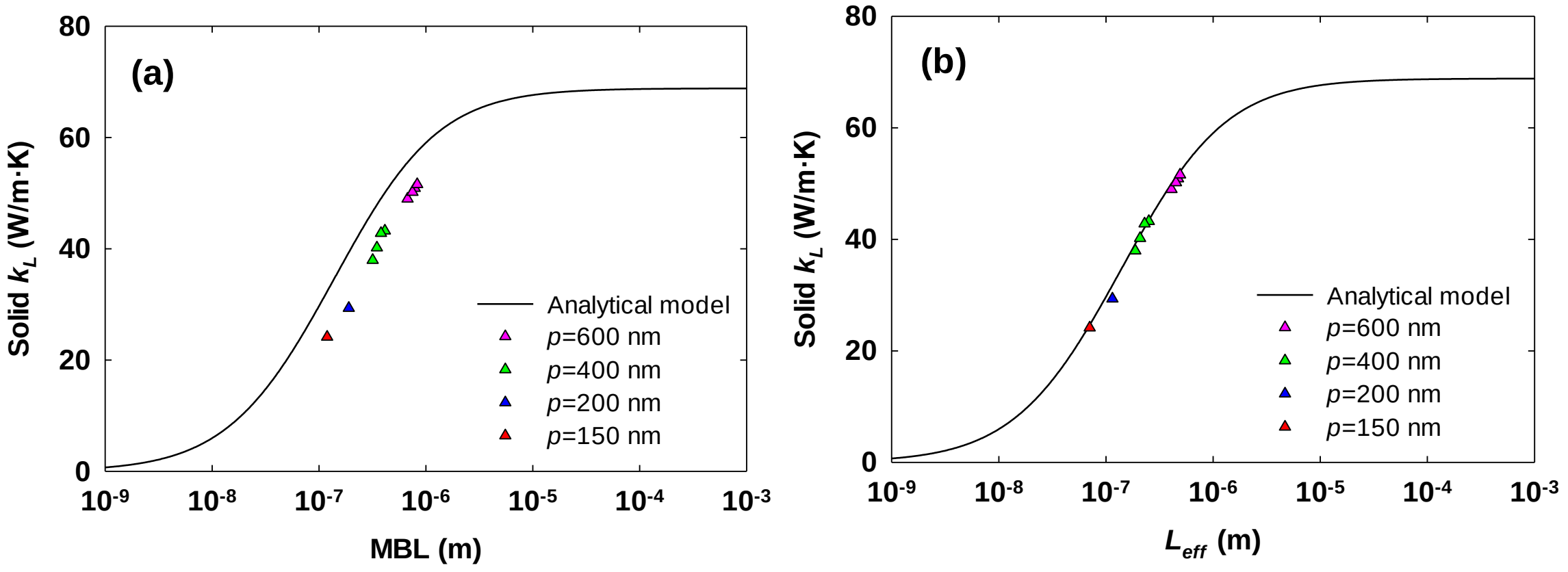

**Fig. 3** Comparison between the prediction by the kinetic relationship using modified phonon MFPs (line) and the phonon MC simulations (symbols) for selected porous patterns across a 220-nm-thick Si film. The employed characteristic length is (a) the MBL or (b) $L_{eff}$ in Fig. 1. Multiple pore diameters are selected for some pitch $p$.

### 2.1.5 Influence of disordered porous patterns

In most calculations, periodic nanopores are assumed. With completely diffusive phonon reflection by pore edges and thus incoherent phonon transport within the structure, calculations of bulk materials with nanopores showed that $k_L$ was insensitive to the spatial configuration and size distribution of the pores.[68] For general bulk materials with randomly embedded nanoparticles, a similar conclusion was reached, in which the volumetric interfacial area $A/V$ was proposed as one key parameter for the thermal conductivity reduction.[69] According to the discussion in Section 2.1.2, however, $A/V(1-\Phi)$ should be used as a more accurate parameter to justify the phonon size effects in the host material. Nevertheless, the spatial distribution of pores should not affect the thermal conductivity when phononic effects are negligible. In fact, the $k_L$ comparison between ordered and disordered nanoporous Si thin films can be used to evaluate the impact of phononic effects.[34, 35, 70] More discussions are given in Section 2.3.

## 2.2 Summary of existing thermal measurements on periodic nanoporous Si thin films

Table 2 summarizes existing measurements of SOI-based nanoporous thin films, as expanded from a summary given by Marconnet et al.[10] In nanofabrication, the porous patterns are first defined by photolithography or EBL,[3, 7, 8, 11-13, 34, 35, 48, 54, 66, 71-73] superlattice nanowire pattern transfer technique,[14] self-assembled block copolymer,[31, 74] or a monolayer film of polystyrene spheres.[31] Nanopores can then be drilled with reactive ion etching (RIE) or deep reactive ion etching (DRIE). Without any mask, a focused ion beam (FIB) is also employed to directly drill nanopores.[2, 66]

Experimentally measured in-plane $k_L$ at 300 K (symbols in Fig. 4) is compared with calculations using $L$=MBL and representative film thicknesses ($t$=22 nm, 220 nm, 2 μm, and ∞) to modify the bulk phonon MFPs (curves in Fig. 4). First-principles-calculated bulk phonon MFPs[26] for Si are employed for the calculations, which has also been validated experimentally.[75] For lightly doped Si, $k \approx k_L$ can be assumed. For heavily doped samples[14, 31, 74], the electronic $k_E$ should be subtracted from $k$ to obtain $k_L$. The Wiedemann-Franz law is employed to compute $k_E = L_0\sigma T$ , in which the metallic Lorenz number $L_0 \approx 2.4 \times 10^{-8}\ W\Omega/K^2$ is approximated for heavily doped samples.[61] All extracted in-plane $k_L$ values are further divided by the Hashin-Shtrikman correction factor in Eq. (3), $F(\Phi) = (1-\Phi)/(1+\Phi)$, to obtain the corresponding solid-film $k_L$. To be consistent, other correction factors used in some cited studies are replaced with the Hashin-Shtrikman factor.

In Fig. 4, some extracted $k_L$ values are significantly lower than the prediction. Despite some early debates on the possible phononic effects within such samples,[1, 2, 4, 6, 7, 14] now it is often acknowledged that phononic effects should be negligible at 300 K for all reported samples.[27, 34, 35, 48] The divergence between measurements and predictions can be attributed to two issues. First, the thermal contact resistance between an employed microdevice and the thin-film sample[2, 14, 31] may lead to an overestimated thermal resistance of a sample and thus an underestimated $k_L$. Possible damage or distortion may also occur during the transfer process of a nanoporous sample. Such issues were addressed in other studies using an integrated device fabricated from the same Si film,[8, 34, 66, 74] micro time-domain thermoreflectance ($\mu$-TDTR) measurements on a suspended sample,[3, 11, 35, 54] and contactless technique of two-laser Raman thermometry.[70] In these studies, the measured $k$ values were mostly comparable to the theoretical predictions at 300 K. Second, more accurate evaluations should also consider the amorphous pore edges introduced by nanofabrication. The significance of pore-edge defects has also been revealed by MD simulations, where non-propagating modes within amorphous pore edges are further considered.[36, 76] In practice, an effectively expanded pore diameter can be used for thermal analysis.[66, 71, 77]

In addition to accurate thermal measurements, attention should also be paid to the nanofabrication techniques used to drill the nanopores. The pore-edge defects, which destroys the phase information of coherent phonons, are directly related to the employed nanofabrication techniques. The commonly used pore-drilling techniques include RIE, DRIE and FIB. The surface damage caused by these techniques and its impact on the transport properties are acknowledged in the past. For instance, $k$ of RIE-patterned Si nanowires (SiNWs)[34] was far lower than that of SiNWs synthesized by the vapor-liquid-solid method.[35] In practice, the effective pore diameter can be justified from the TEM[31, 66] or scanning electron microscopy (SEM)[12, 35] studies. The outer edge of the pore can be treated as the effective pore size. The thermal conductivity reduction due to interface/surface amorphization can also be found in a Si nanobeam with deposited Al nanopillars, where the reduced thermal conductivity mainly results from the phonon scattering at the pillar/beam interface with intermixed aluminum and silicon atoms.[78] George et al. further demonstrated a 30–40% lower thermal conductivity in silicon membranes covered with aluminum films, leading to significantly enhanced thermoelectric performance.[79]

Figure 5 shows the measured in-plane $k_L$ of four DRIE-drilled samples and two FIB-drilled samples.[66] Based on the TEM and SEM studies, the pore radius is expanded by 13–40 nm for DRIE samples with pore diameters of 94 to 300 nm, and ~50 nm for FIB samples with pore diameters of 200 and 300 nm. Using such effective pore diameters, $k_L$ values predicted by phonon MC simulations agree well with the measurement results over the whole temperature range.

**Table 2.** Thermal measurements of nanoporous Si films. A square (Sq.), staggered (Sta.) or hexagonal (Hex.) lattice for the periodic porous pattern is often used. One paper further compares nanoporous films made of single-crystal and polycrystalline (poly) Si.[72]

| Article | Year | Measurement configuration | Doping | $T$ (K) | Pattern | $t$ (nm) | $d$ (nm) | $p$ (nm) |
|---|---|---|---|---|---|---|---|---|
| Hao et al.[66] | 2018 | $3\omega$ technique for a suspended heater bridge | Undoped | 80–300 | Sq. | 220 | 120–400 | 150–600 |
| Anufriev et al.[80] | 2017 | $\mu$-TDTR | Undoped | 30–296 | Sq., Sta. | 145 | 80–425 | 160–500 |
| Graczykowski et al.[70] | 2017 | Two-laser Raman thermometry | Undoped | 300–900 | Sq. | 250 | 130–140 | 200–300 |
| Lee et al.[34] | 2017 | Suspended heater-thermometer | $p$-type doped: $10^{15}/cm^3$ | 14–325 | Aligned rectangular holes [a)] | 80±10 | 60±5 | 80–120 |
| Maire et al.[35] | 2017 | $\mu$-TDTR | Boron-doped: $10^{15}/cm^3$ | 4–18, 300 | Sq., Disordered [b)] | 145 | 133 and 161 | 300 |
| Verdier et al.[12] | 2017 | $\mu$-TDTR | Undoped | 4, 300 | Sq., Sta. | 145 | 90–425 | 200–500 |
| Anufriev et al.[54] | 2016 | $\mu$-TDTR | Undoped | 4, 300 | Sq., Hex., Honeycomb | 80 | 65–240 | 120–280 |
| Nomura et al.[11] | 2016 | $\mu$-TDTR | Boron-doped | 300 | Sq., Hex. | 145 | 120–275 | 300 |
| Wagner et al.[13] | 2016 | Two-laser Raman thermometry | Undoped | 300 | Sq., Disordered | 250 | 175 | 300 |
| Nomura et al.[72] | 2015 | $\mu$-TDTR | Boron for single-crystal Si; undoped poly Si | 300 | Sq. | 145, 143 | 74–254 (poly Si), 112–270 (single-crystal Si) | 300 |
| Alaie et al.[2] | 2015 | Suspended island technique | Undoped | 300 | Sq. w/ smaller interpenetrating holes | 366 | 850 | 1100 |
| Lim et al.[74] | 2015 | Suspended heater-thermometer | Boron-doped: $3.1\times10^{18}/cm^3$ –$6.5\times10^{19}/cm^3$ | ~12 – ~320 | Hex. | 100 | 26–44 | ~60 |
| Kim et al.[8] | 2012 | Suspended bridge | *p*-type boron: $10^{16}/cm^3$ | 300 | Sq. | 500 | 204–525 | 500–900 |
| Marconnet et al.[48] | 2012 | Suspended heater bridge | Undoped | 300 | Sq. [c)] | 196 | 110–280 | 385 |
| Hopkins et al.[7] | 2010 | TDTR (cross-plane measurement) | Electrical resistivity as 37.5–62.5 Ω cm | 300 | Sq. | 500 | 300 and 400 | 500–800 |
| Tang et al.[31] | 2010 | Suspended heater-thermometer | Intrinsic: $3\times10^{14}/cm^3$, and boron-doped: $5\times10^{19}/cm^3$ | 25–300 | Hex. | 100 | 32–198 | 55–350 |
| Yu et al.[14] | 2010 | Suspended heater-thermometer | Boron-doped: $2\times10^{19}/cm^3$ | 90–310 | Sq. (circular and square holes) | 22 | 11, 16 and 270 | 34, 385 |
| Song et al.[71] | 2004 | Suspended film (steady-state Joule heating) | *n*-type: $5\times10^{14}$ –$5\times10^{15}/cm^3$ | 50–300 | Sq., Sta. | 4450– 7440 | 1900, 2300 and 10900 | 4000 and 20000 |

a) Structure includes periodic and aperiodic rectangular holes.
b) 1D (single row of holes) and 2D patterns.
c) 1D single row pattern in the middle of the nanobeams. The porosity is only for the porous region.

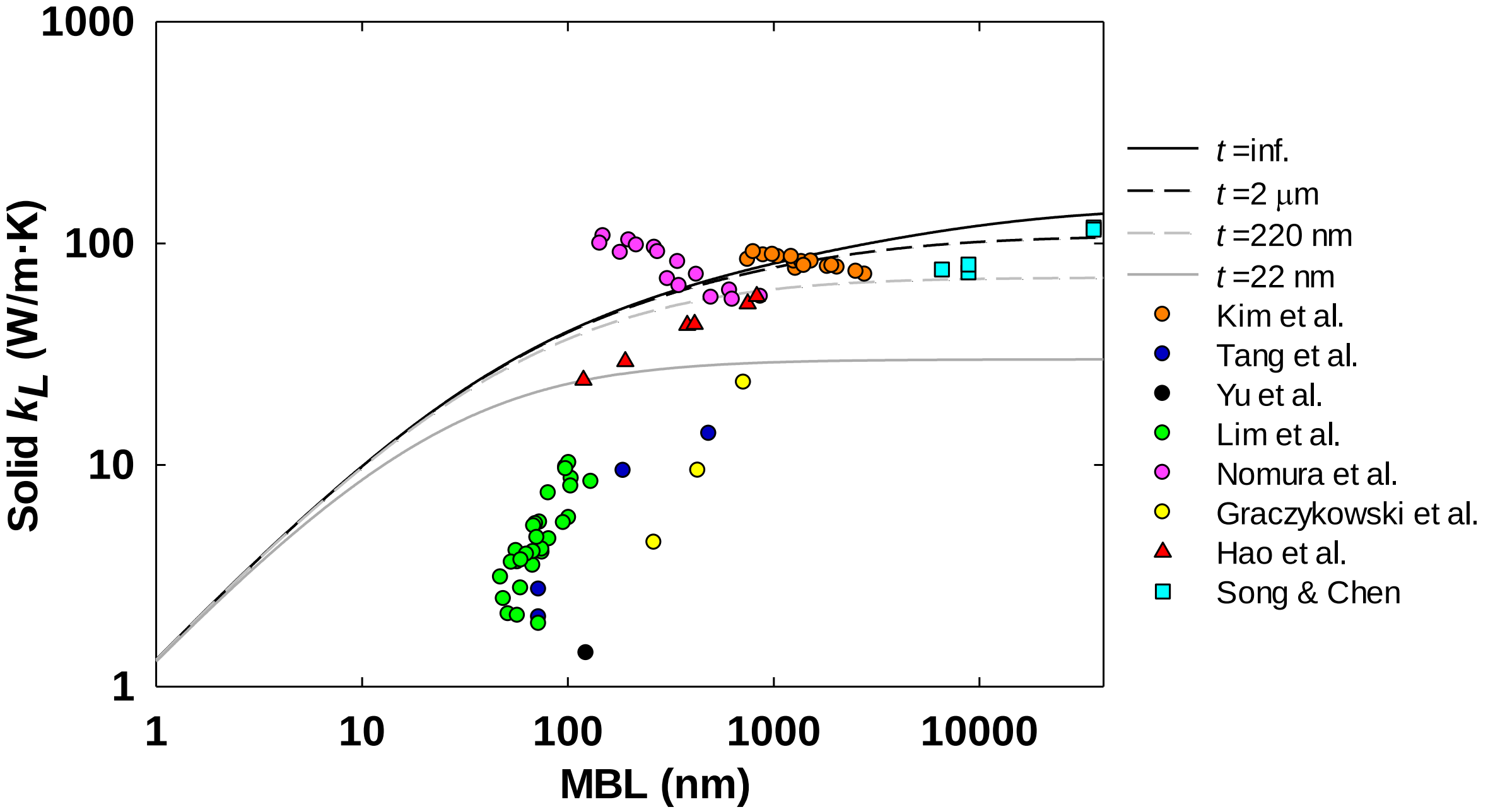

**Fig. 4** Comparison between predicted (lines) and measured (symbols) in-plane $k_L$ of porous Si films at 300 K. With $F(\Phi)$ corrections, all values are for the corresponding solid film.

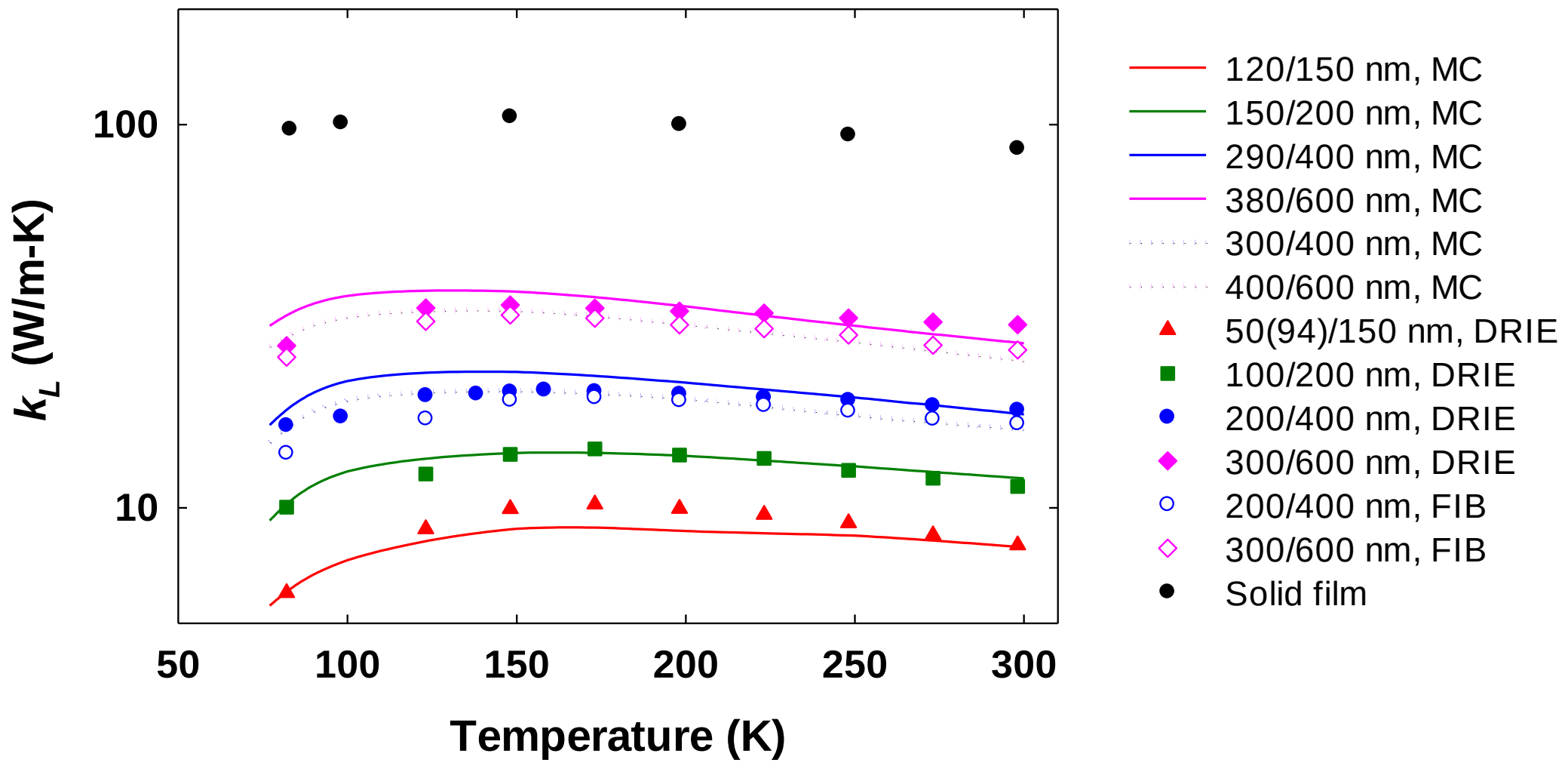

**Fig. 5** Temperature-dependent in-plane $k_L$ of the solid and nanoporous Si thin film (symbols), in comparison to predictions by MC simulations (lines with the same color) using an effective pore diameter indicated in the legend. The diameter/pitch combinations are given in the legend. The solid lines are for DRIE samples and the dashed lines are for two FIB samples. No correction factor is used to normalize $k_L$. Reproduced from Ref. [66], with Creative Commons Licenses.

### 2.3 Justification of phononic effects within periodic nanoporous thin films

In practice, phononic effects for phonons are hard to be justified by comparing measurements with predictions assuming completely diffusive pore-edge scattering. This is due to the uncertainties in $k_L$ predictions due to the pore-edge defects and employed frequency-dependent bulk phonon MFPs. In this section, other experimental methods to justify the phononic effects are discussed in brief.

The first method is based on $k_L$ comparison between periodic and aperiodic nanoporous films.[13, 34, 35, 70] By comparing $k_L$ for periodic and aperiodic samples, phononic effects were observed only below 10 K for nanoporous Si films with $p$=300 nm,[35] and below 14 K for Si nanomeshes with $p$>100 nm.[34] At elevated temperatures, the reduced phonon wavelengths require smaller structures and smoother pore edges so that phononic effects becomes negligible. This finding was consistent with the measurements on Si nanoporous films with $p$ of 200–300 nm, where incoherent phonon transport was found above 300 K.[70]

The second method is to check the possible variation of the phonon dispersion. In one study by Graczykowski et al., the phonon dispersion of nanoporous Si films was measured with Brillouin light scattering.[81] Along another line, the specific heat of the nanoporous film, as determined by the phonon dispersion, can also be compared to that for bulk Si.[66] A similar heat capacity comparison has also be used to justify the possible phonon dispersion modification for a Si wafer with patterned periodic nanopillars.[82] In physics, strong phononic effects should lead to strong variation in the phonon specific heat that solely depends on the phonon dispersion.[83] In contrast, the thermal conductivity is affected by both the phononic effects and pore-edge defects. In early experimental studies,[2, 3, 7, 8, 11-14, 31, 34, 35, 48, 54, 66, 71-74, 84] only the thermal conductivity is measured and no information is available for the specific heat. In this aspect, both the specific heat and thermal conductivity can be measured simultaneously for the same sample using the $3\omega$ technique for a suspended sample,[85] providing new insights into the possible coherent phonon transport. Figure 6a and 6b present the solid volumetric specific heat $C$ for all bilayer films patterned with DRIE and a FIB, respectively. All measured $C$ values are divided by $(1-\Phi)$. In general, the solid $C$ values of nanoporous Si films follow that for a solid film, i.e., a film without nanopores (black dots in Fig. 6a). Some divergence can be attributed to the inaccuracy in $\Phi$ and other defects, considering the wavy pore sidewalls for DRIE samples and tapered sidewalls for FIB samples.

At ultralow temperatures, the bulk phonon MFPs are much longer than $L$ so that $\Lambda_{eff} \approx L$ can be assumed. In this situation, the temperature dependence of $k_L \approx C v_g L/3$ is mainly determined by $C$ and thus the phonon dispersion.[52] Instead of measuring $C$ directly, the comparison of the power law $k_L \sim T^n$ also indicates the possible change in the phonon dispersion.[18] In the work by Lee et al.[34], a clear $k_L \sim T^3$ trend is observed at low temperature, indicating bulk-like temperature dependence of the specific heat and thus no phononic effects. As one major restriction, this approach is not applicable at 300 K or above, at which $\Lambda_{eff}$ and thus $k_L$ are further reduced by the temperature-dependent internal phonon-phonon scattering. In this situation, the temperature dependence of $k_L$ does not follow the trend of the specific heat $C$.

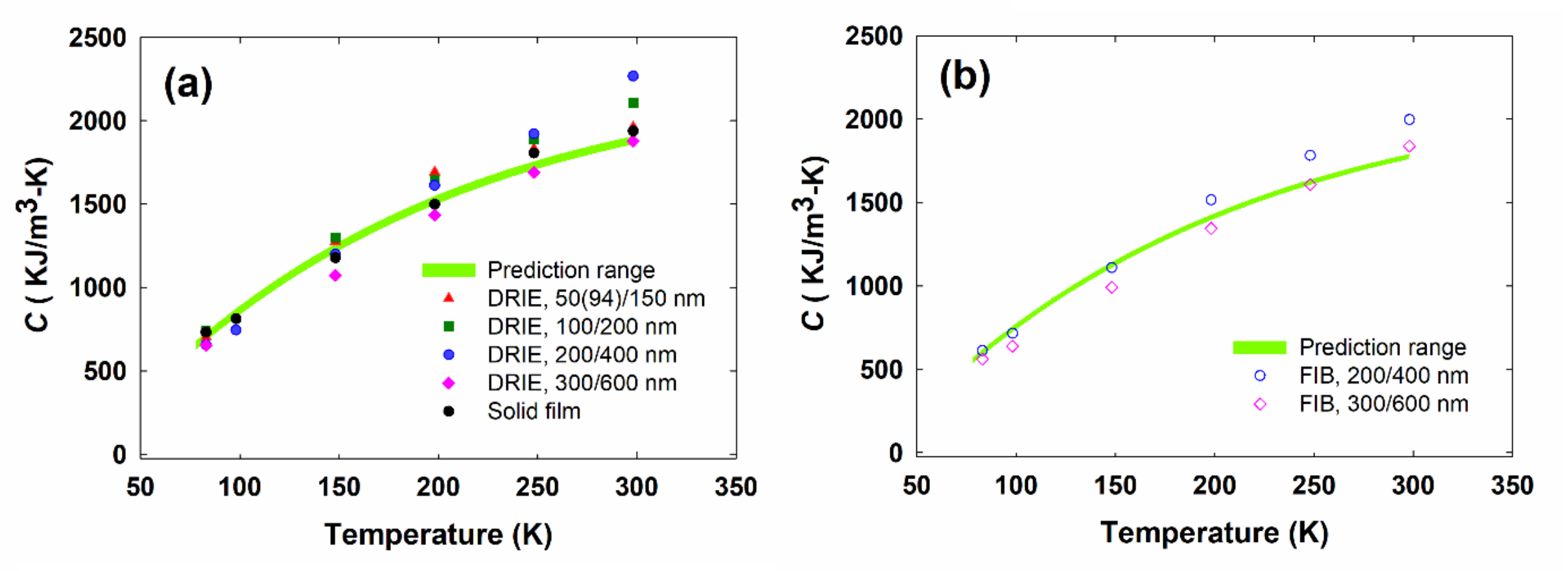

**Fig. 6** Temperature-dependent solid *C* of bilayer films drilled by (a) DRIE and (b) a FIB, in comparison to the prediction using bulk *C* values for metals and Si. Reproduced from Ref. [[66]], with Creative Commons Licenses.

## 2.4 Direct porous film growth to minimize pore-edge defects

As discussed above, the pore-edge defects associated with the pore-drilling processes can add uncertainties to the phonon transport analysis. When coherent phonon transport is desired, smooth pore edges are critical to the conservation of coherent phonon phases. One way to minimize pore-edge defects is to directly grow nanoporous films with MOCVD, instead of drilling pores after the thin film growth. To block the local growth in porous regions, an array of vertical $SiO_2$ nanopillars can be fabricated on a sapphire substrate as masks. After the film growth at high temperatures, these $SiO_2$ nanopillars can then be etched off with hydrogen fluoride to yield a nanoporous thin film. Room-temperature phononic effects can be possibly observed if ultrafine nanoporous patterns can be fabricated with this approach.

Figure 7 shows the measured cross-plane $k$ of nanoporous multilayered films at 300 K, in comparison to theoretical predictions using the MBL and film thicknesses to modify the bulk phonon MFPs within each layer.[39] Incoherent phonon transport is assumed in the calculations. Each film consists of three layers, i.e., a 50-nm-thick low-temperature GaN nucleation layer, a 50-nm-thick GaN buffer layer, and a 150-nm-thick $In_{0.1}Ga_{0.9}N$ layer. The SEM images of representative samples are shown in the inset, with a fixed 300 nm pore diameter for all fabricated patterns. The measured $k$ via the TDTR method generally agrees well with the theoretical predictions, with some measurement errors due to the thermal penetration into the substrate. The conclusion here is in contrast with the cross-plane thermal studies of nanoporous Si films by Hopkins et al., where phononic effects have been suggested to play an important role in the $k$ reduction.[7] However, the $k$ values of nanoporous films are anticipated to be underestimated by Hopkins et al. because their calibration of a solid Si film also yields $k$ much lower than theoretical predictions.[86, 87]

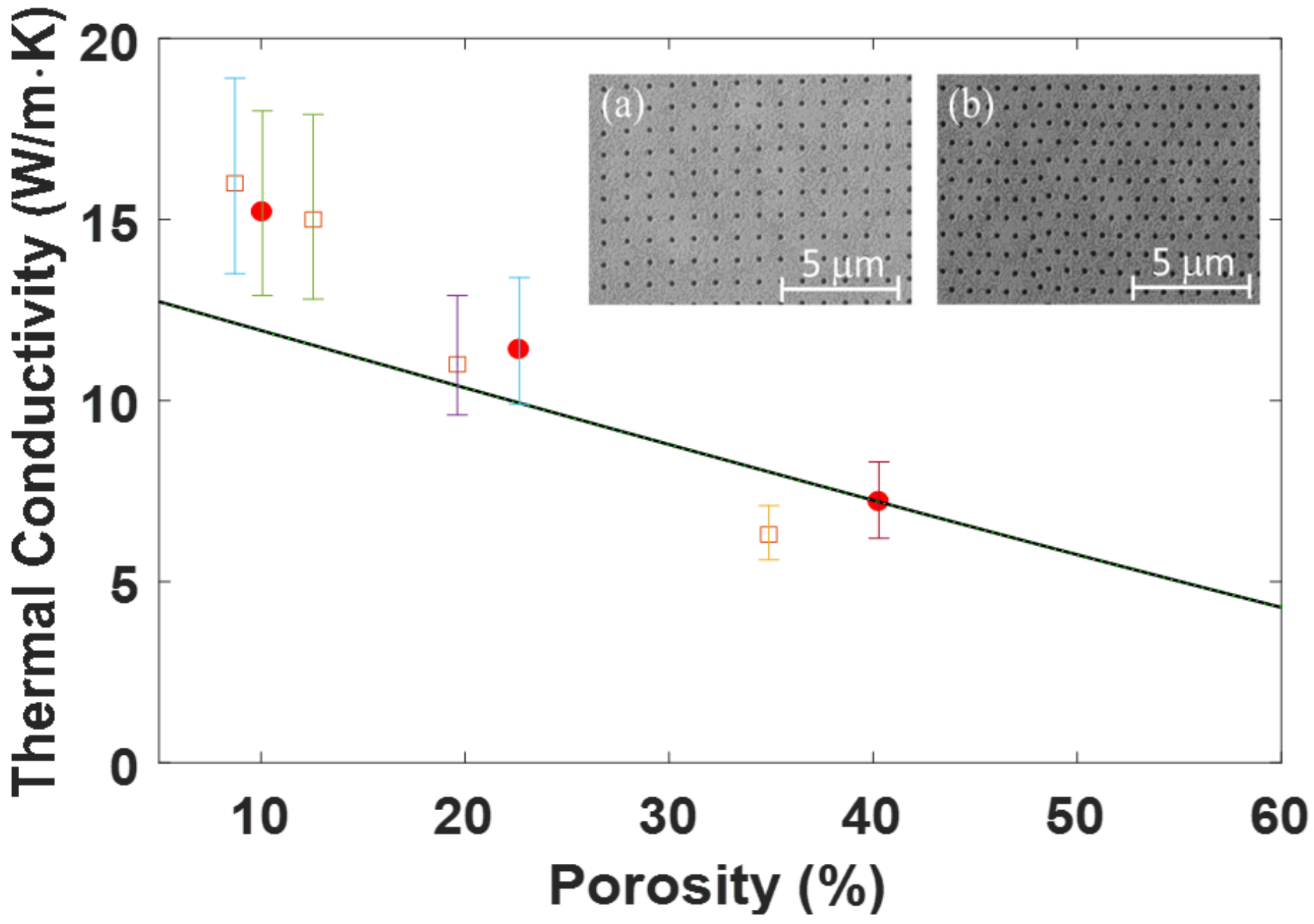

**Fig. 7** Comparison between the measured and predicted $k$ values for tri-layered nanoporous GaN-based films. Here filled circles are for hexagonal patterns, whereas empty squares are for patterns on a square lattice. The insets show the SEM images for representative samples. Reproduced from Ref. [[39]]. Copyright 2018 Springer Nature.

**3. Thermal studies of antidot lattices (2D phononic crystal)**

With phonon transport is completely constricted within an atomic-thick sheet, periodic porous 2D materials are also studied in recent years though nanofabricated feature sizes are still much longer than the room-temperature phonon wavelength to induce strong phononic effects at 300 K. Among these materials, GALs have received enormous attention as an effective way to modify the intrinsic transport and optical properties of pristine graphene for applications in electronic/optoelectronic devices,[88-90] waveguides,[91] and thermoelectrics.[19, 92, 93] For thermal applications, GALs also belong to "graphene phononic crystals" (GPnCs, shown in Fig.11a), similar to periodic nanoporous Si films. Compared with thin films with through-film pores, finer porous patterns can be possibly achieved in 2D materials because the maximum aspect ratio for dry etching does not pose any restriction for atomically thin layers. Numerous calculations have been carried out on GALs with a pitch $p$~1 nm and pore diameter $d$<1 nm. For instance, an up to unity room-temperature thermoelectric figure of merit is predicted by Yan et al.[92] but no experimental validation is available now. In the future, such atomically porous structures can be achieved with advancement in nanofabrication techniques. For graphene, the exact pore-edge configuration, namely armchair and zigzag edges,[94] is anticipated to affect the phonon scattering as well. The rich physics associated with GALs and other 2D antidot lattices provides new opportunities in exploring phonon transport in a periodic material. Similar thermoelectric studies have also been carried out by Lee et al. on nanoporous Si thin films with the pore size and neck width both in the range of 0.63−2.26 nm.[95, 96] Experimental validation is unavailable for now due to the challenge in fabricating such ultrafine structures.

Experimentally, phononic studies of GALs have major challenges in the nanofabrication of periodic atomic-level pores and accurate thermal measurements. Because $k$ can be largely suppressed by a substrate,[97] all thermal measurements may be carried out by suspending such fragile materials. The samples can be easily damaged and distorted during the process. In the literature, the only thermal measurement on graphene nanomeshes was carried out by Oh et al. with a hexagonal pattern.[98] With a pitch of tens of nanometers, irregular pores were etched across graphene, instead of a highly ordered pattern often assumed in calculations. Using a suspended microdevice, other porous graphene was measured by drilling pores with a FIB across graphene.[99, 100] In general, above nanoporous patterns, with ~10 nm or sub-micron pitches, are still too large to compare with superlattices with down to sub-10 nm periodic lengths. Furthermore, highly amorphous pore edges are expected for FIB-drilled patterns, such as an amorphous surface layer of ~10 nm depth, lattice defects (vacancies, interstitials, and dislocations), Ga ion implantation, and large atom displacement within the collision cascade that extends tens of nanometers from the cut surface.[101] In this aspect, scanning helium ion lithography can be better for the nanofabrication of 2D materials.[102] For a suspended graphene sample, sub-1 nm pores can be directly patterned with focused electron beams under a TEM.[103] However, the high-energy electrons also cause defects or structural changes, shown as a concentric ring-like structure around pores. For phononic studies, these pore-edge defects must be considered.

### 3.1 MBL of 2D nanoporous materials

Different from thin films, the phonon boundary scattering by the top and bottom surfaces is eliminated in suspended 2D materials to simplify the analysis. For GALs, ballistic electron transport is suggested because the structure size is often much shorter than majority electron MFPs in pristine graphene.[104, 105] The same conclusion may also be anticipated for phonons in graphene so that $k_L$ can be computed assuming $\Lambda_{eff} \approx MBL$ for all phonons. This argument is known as the small-nanostructure-size limit, where $\Lambda_{eff}$ approaches the characteristic length of nanostructures.[106, 107] The existing derivation of the MBL for a 3D enclosure[108] can be extended to a 2D enclosure and used for phonon modeling. Due to the decreased degrees of freedom for phonon movement, phonons travel a shorter distance between their successive collisions with pore edges, leading to a shorter 2D MBL than quasi-2D MBLs given in Eq. (4). In this review, the existing MBL derivation for a 3D enclosure[108] is extended to a pure 2D enclosure.

Within a plane, consider an isothermal and nonscattering medium that is surrounded by a 2D blackbody enclosure with perimeter $P$ (Fig. 8a). In such an enclosure, the spectral heat flux onto any point on the wall is given by the integration of incident directional intensity $I_{b\eta}$, where the subscript $b$ indicates blackbody and $\eta$ is the wavenumber. The integration is over the hemispherical solid angle $\Omega = 2\pi$ for a 3D enclosure, and is over a half-circle solid angle $\Omega = \pi$ for a 2D enclosure. The wavenumber-dependent heat flux $q_\eta$ follows the same expression as a 3D enclosure[108] except for the change in the solid angle integration:

$$
\begin{aligned}
q_\eta &= I_{b\eta} \int_{\text{Half circle}} \left(1 - e^{-\kappa_\eta L_d(\hat{s})}\right) cos\theta d\Omega \\
&= 2I_{b\eta} \int_{\theta=0}^{\frac{\pi}{2}} (1 - e^{-\kappa_\eta L}) cos\theta d\theta \\
&= 2I_{b\eta}(1 - e^{-\kappa_\eta L}),
\end{aligned} \tag{6}
$$

in which an effective length $L$ as the MBL replaces the direction-dependent distance $L_d(\hat{s})$ from this point to the wall along the $\hat{s}$ direction (Fig. 8b). When $\kappa_\eta L \to 0$ for a optically thin medium, Eq. (6) becomes

$$q_\eta \approx 2I_{b\eta}\kappa_\eta L. \quad (7)$$

On the other hand, total radiation received by the wall along its perimeter, $q_\eta P$, is from the blackbody emission inside the whole medium-filled area $S$ within the 2D enclosure. For an optically thin medium, no self-absorption occurs and $q_\eta P$ is given as

$$q_\eta P = (2\pi I_{b\eta})S\kappa_\eta. \quad (8)$$

In the derivation of 3D MBLs, $q_\eta A = (4\pi I_{b\eta})V\kappa_\eta$ is instead suggested for a 3D enclosure with inner wall surface area $A$ and filled volume $V$.[108] Comparing Eqs. (7) and (8) gives the effective MBL as

$$L = \frac{\pi S}{P}. \quad (9)$$

For nanoporous 2D materials, $S$ corresponds to the solid area $S_{Solid}$ of a period and $P$ is the pore perimeter. By fitting the MCRT results for a 2D material with aligned pores, $L \approx 3.4S_{Solid}/P$ is obtained and is close to the $L$ expression in Eq. (9).[42]

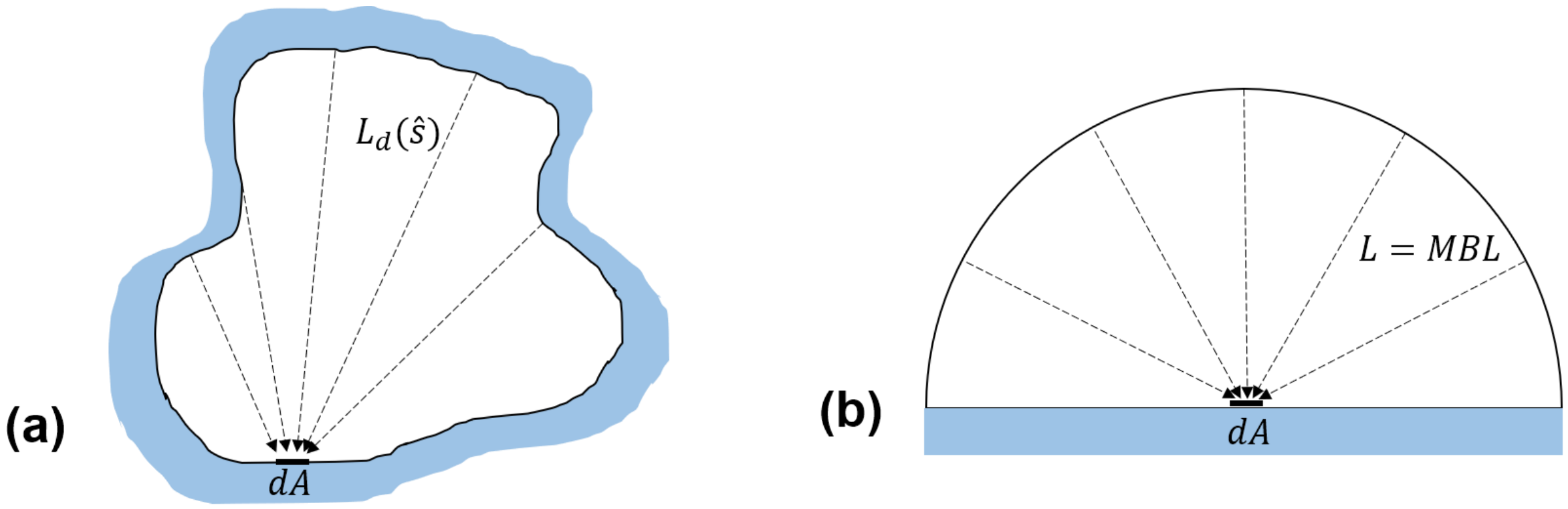

**Fig. 8** (a) An arbitrary 2D enclosure filled with an isothermal and nonscattering medium. (b) Equivalent half circle radiating to the center of its base as the receiving point in (a).

### 3.2 Existing thermal studies of GALs

Currently, measurements on GALs are focused on their electrical properties.[53, 88, 104, 105, 109] Instead of GALs with aligned pores, thermal measurements are only available on graphene nanomeshes with irregular pores.[98] This study employed the optothermal Raman technique, where a Raman laser was used to heat up a sample suspended across a hole and the associated temperature rise was read from the Raman peak shift.[110, 111] The effective $k$ of a GAL can thus be obtained from 2D heat conduction analysis. Such measurements often have large uncertainties due to 1) the difficulty in determining the actual laser power absorbed by a sample and 2) complexity introduced by the strong thermal nonequilibrium between electrons, optical phonons and acoustic phonons inside graphene.[112] More comparison measurements with accurate measurement techniques should be carried out to better understand the coherent and incoherent phonon transport within periodic nanoporous structures.

Different from real samples with inevitable defects, atomistic simulations assuming periodic atomic structures often suggest much stronger phononic effects. In simulations and modeling, it is found that the thermal conductivity of a GAL can be significantly lower than that of graphene and can be efficiently tuned by changing the porosity and period length.[115] Further phonon mode analysis reveal that the thermal conductivity reduction is due to the increasing of phonon localization and three-phonon scattering in a GAL.[113]

### 3.3 Extension to nanopillar-modified graphene and thermal rectification applications

Instead of drilling pores, introducing periodic pillars on two sides of graphene nano-ribbon to form graphene nano-ribbon with nano cross junctions (NCJs) (Fig.9b) will induce phonon local resonant hybridization, which also greatly reduces the thermal conductivity. As another type of phononic crystals, such pillar-modified structures were first proposed by Davis et al. for Si films.[114] Extensive studies can be found elsewhere.[82, 115-120] As an interesting result in simulations, the thermal conductivity of graphene nano-ribbon with NCJs increases after replacing $C^{12}$ in pillars with isotopic atoms (lighter or heavier than $C^{12}$). It is caused by the mismatch between resonant modes and propagating modes, which breaks and decreases the original hybridization and facilitates phonon transport.[121]

In addition to the decreased thermal conductivity, the temperature dependence of thermal conductivity for graphene can also be modulated. Compared to pristine graphene, the thermal conductivity of GALs (Fig. 9a) has a weaker dependence on temperature, $k \sim T^{-\alpha}$. The power exponent ($\alpha$) can be efficiently tuned by changing the characteristic size of such phononic crystals.[122]

A potential application of GAL is for thermal rectification. It is found that the asymmetric graphene/GAL system is a promising thermal rectifier[123], because of the different temperature dependence of thermal conductivities in graphene and GAL. It is also observed in experiments.[101] A thermal rectification factor of 26% is achieved in a defect-engineered monolayer graphene with nanopores on one side.[99] This remarkable thermal rectification results from the difference in temperature dependence of $k$ for graphene and GAL regions, which can be regarded as nonlinearity introduced into the system.[124] Anticipated to be weak around 300 K for FIB-drilled graphene,[125] phononic effects are not required to achieve this difference. To enhance the thermal rectification ratio, a series circuit of thermal rectifiers is proposed.[123] Hu et al. studied thermal rectification in both two-section and three-section asymmetric graphene/GAL structures. Similar to the series effect in electronic circuits, the series effect is demonstrated by the consistence between the results of theoretical prediction and that of MD simulations. It is found that both the series effect and size effect are effective strategies to enhance the thermal rectification ratio.

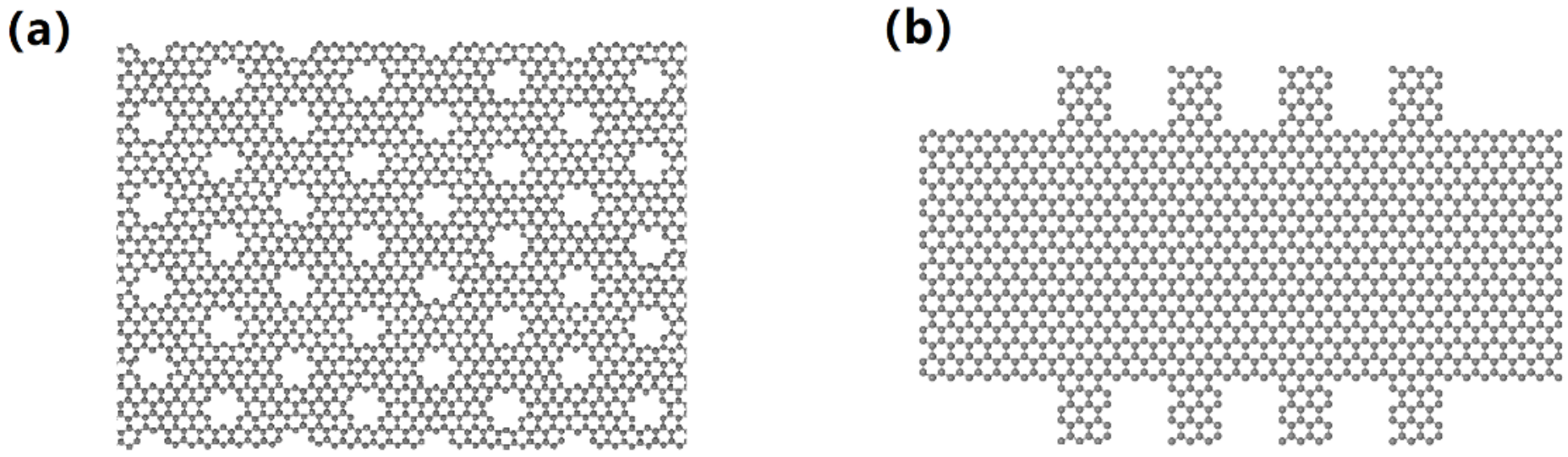

**Fig. 9** The structure of a (a) GAL with hexagonal porous patterns, (b) graphene nano-ribbon with NCJs.

## 4. Porous nano-cages (3D phononic crystal)

3D periodic materials have attracted great attention due to their potential application in thermoelectrics. Compared to 1D and 2D structures, a 3D phononic crystal is bulk and are more suitable for common applications. Periodic spherical porous 3D Si phononic crystal (Fig. 10a), SiNW-cage structure (Fig. 10b) and isotopic 3D Si phononic crystal (Fig. 10c) are studied. The Si phononic crystal is constructed by periodic arrangement of nanoscale supercells constructed from a cubic cell with a spherical pore. The SiNW-cage consists of SiNWs connected by NCJs. The isotopic 3D Si phononic crystal is assembled periodically in three directions by $^{28}$Si and $^{M}$Si atoms.

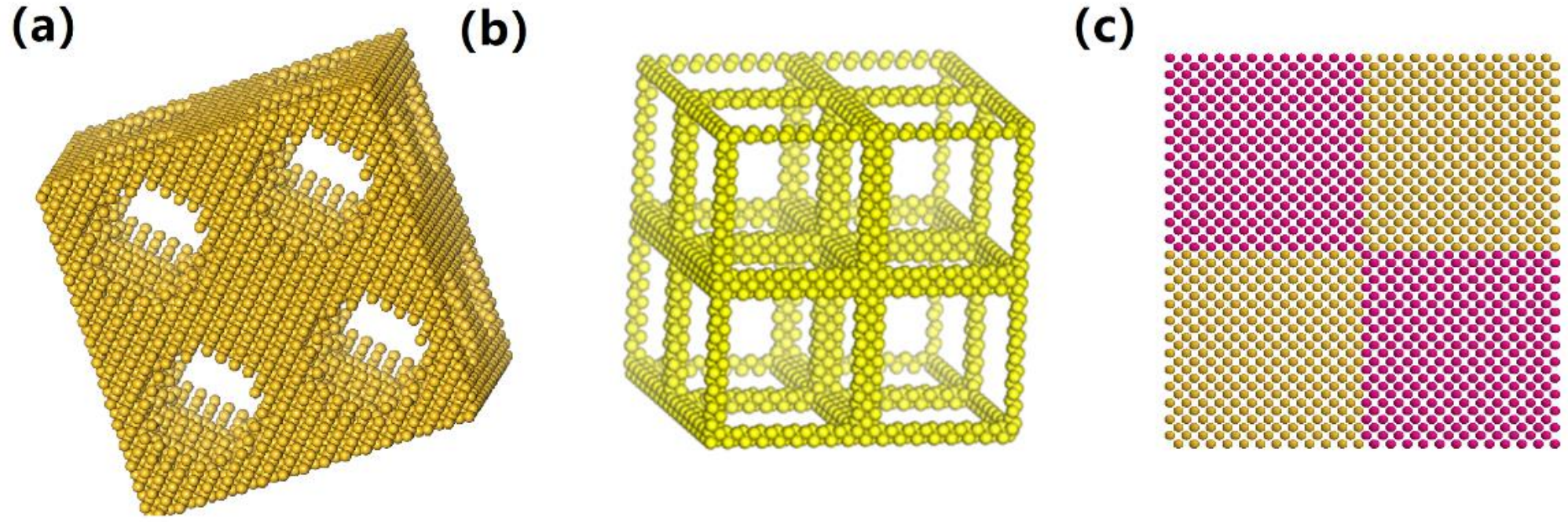

**Fig. 10** The structure of a (a) periodic spherical porous 3D phononic crystal, (b) SiNW-cage and (c) isotopic 3D Si phononic crystal.

### 4.1 Computed thermal conductivities of 3D phononic crystals

Phonon wave effects can effectively reduce the thermal conductivity of a Si phononic crystal, SiNW-cage structure and isotopic 3D Si phononic crystal. It is found that the thermal conductivity of a Si phononic crystal is decreased to 0.022 W/m·K, which is only 0.01% of the thermal conductivity for bulk Si. This low thermal conductivity decreases as the porosity increases. The reduction in the thermal conductivity is due to more phonons localized in a Si phononic crystal at boundaries.[126] In contrast to the huge thermal conductivity reduction, the electronic transport coefficients of a Si phononic crystal at 300 K are reduced slightly, and the Seebeck coefficient is similar to that of bulk Si. This leads to a higher ZT = 0.66 of a Si phononic crystal, which is about 66 times of that of bulk Si.[127] For SiNW-cage structures, the thermal conductivity can be as low as 0.173 W/m·K , which is even one order of magnitude lower than that of SiNW.[128] The large reduction in thermal conductivity is due to significant phonon local resonant hybridization at the junction part which reduces group velocity in a wide range of phonon modes.[128] The mechanism here are not excepted to scatter electrons, which will increase the ZT. For an isotopic 3D Si phononic crystal, the thermal conductivity is significantly reduced at 1000 K. Increasing the mass ratio will further decrease the thermal conductivity. The decrease of thermal conductivity in a isotopic 3D Si phononic crystal is attributed to both the decrease of group velocities and the localization.[129]

### 4.2 Quantifying phonon particle and wave transport in 3D phononic crystals

In periodic structures with smooth boundaries or interfaces, the mix of wave and particle phonon transport is expected though phonon wave transport dominates. During the past decades, many researches have focused on quantitative understanding of particle and wave transport of electrons and photons.[130, 131] In order to optimally manipulate phonon transport, the direct individual contributions of phonon particle and wave effects to the modulation of thermal conductivities need to be evaluated.

Phonon particle and wave effects can be quantified by combining MC and atomic Green's function (AGF) methods.[132] To probe phonon transport from the particle standpoint, the MC method can be used for a NCJ system. The transmittance can be used to decide whether phonons can transport across the junction part (Fig. 11a). The phonon wave information is included in this transmittance. To incorporate the transmittance into the MC simulation, a random number is drawn from a uniform distribution for every phonon. By comparing this random number with the transmittance, we decide whether the phonon can transport across the junction part or not. Here, the AGF method is used to obtain the transmittance. The combination of the AGF and MC methods is termed as AGFMC. As shown in Fig. 11b, with the introduction of cross junction, $k_{NCJ_MC}$ (blue dot) is smaller than $k_{SiNW}$ (black dot). This is because the junction increases the phonon scattering. In addition, $k_{NCJ_AGFMC}$ (red dot) is even smaller than $k_{NCJ_MC}$ (blue dot). This is due to the enhanced blockage originating from phonon resonant hybridization as wave effects, which has been further incorporated into the AGFMC. The fraction of thermal conductivity reduction by wave effects to the total thermal conductivity reduction is shown in Fig. 11c. When the cross section area (CSA) increases from 2.23 nm$^2$ to 17.72 nm$^2$, the wave ratio ($\eta_{wave}$), which measures the fraction of thermal conductivity reduction by the wave effect to the total thermal conductivity reduction, decreases monotonously. This shows that the wave effect weakens as the system size increases, which is in accordance to previous studies. What is more striking, as shown in Fig. 11c, the wave effect is only 68% (corresponding to 32% particle effects), even for the CSA as small as 2.23 nm$^2$ for 4-leg NCJs (shown in the inset of Fig.11c). This accentuates the importance of mutually controlling phonon particle and wave characteristics in NCJ.

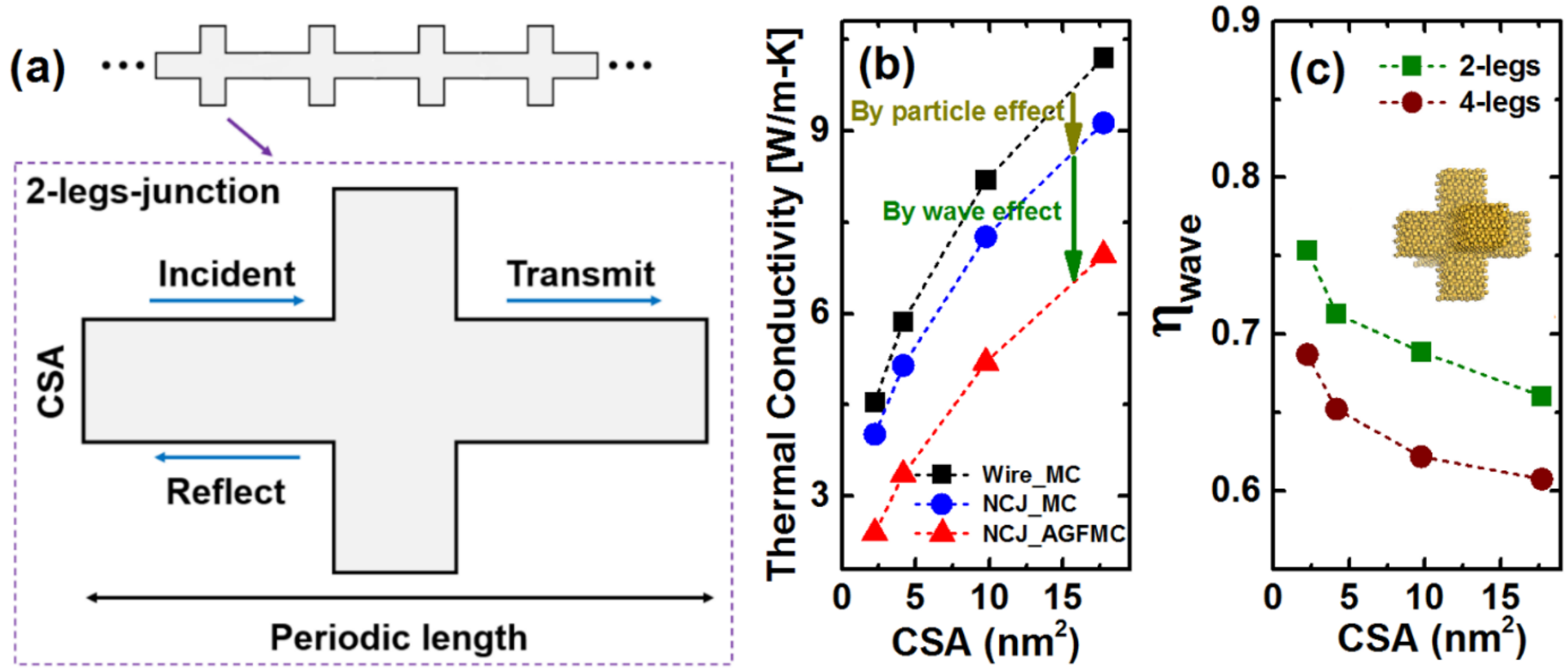

**Fig. 11** (a) Schematic picture of phonon transport across a NCJ system. (b) The thermal conductivity of a SiNW (black dot) and NCJ (red and blue dot) as a function of the CSA at 300 K. The data of the red line are obtained by MC, which only takes the phonon particle effect into account. The data of the blue line are obtained by the AGFMC, which considers both phonon particle and wave effects. (c) The ratio of the thermal conductivity reduction by the phonon wave effect to the total thermal conductivity reduction versus the CSA.

**4.3 Simulations to incorporate both wave and particle transport**

In periodic nanostructures, wave effects are anticipated for some long-wavelength phonons and particle view should be valid for phonons with very short wavelengths. When wave effects become notable with a reduced periodic length and/or at cryogenic temperatures, the two phenomena can coexist. Such particle and wave phonon conduction can be quantified by a two-phonon model,[133, 134] which can be used for quantifying wave conduction in periodic phononic structures. But the two phonon model considers all phonons as two gray media. The fact that different phonon modes can have different wavelengths is significantly neglected, and the non-equilibrium between phonons are not considered.[135] In earlier studies on superlattices, a modified lattice dynamics model has been proposed, in which incoherent interfacial phonon scattering is incorporated with a complex wave vector involving the phonon MFP.[28, 136] In one recent work, an improved phonon MC technique has been proposed, where a particle model can further include wave properties to model interferences between wave packets.[137] This technique is suitable for periodic structures with length scales larger than phonon coherence length (e.g., 1 nm at 300 K for Si[61]), where coherent phonon transport can still take place.[138]

**5. Summary and Perspective**

Similar to photonic crystals, phononic crystals based on periodic structures provide new opportunities in manipulating lattice vibrational waves. To have a strong influence on heat conduction by wave effects, two conditions must be satisfied: 1) the pitch should be comparable to the majority phonon wavelengths but much shorter than phonon MFPs; 2) the interface/surface roughness should be smaller than the phonon wavelength to minimize the diffusive phonon scattering that destroys the phase information of phonons. The influence of more complicated mechanisms, such as the Akhiezer mechanism,[15] may also be considered in detailed studies. This mechanism is due to the coupling of the strain of sound waves and thermal phonons. In practice, the Akhiezer mechanism can largely suppress the contribution of low-frequency phonons that are more likely to be affected by the phononic effects.[15]

In Si, majority phonon wavelengths at 300 K are of 1–10 nm.[26, 27, 139] Therefore, the reported room-temperature phononic effects[2, 7] in nanoporous structures can be mostly attributed to measurement errors and other structural defects such as amorphous pore edges. Phononic effects only become critical at around 10 K or below for nanoporous Si films, when heat conduction is dominated by phonons with wavelengths of tens of nanometers. However, such ultra-low-temperature phononic crystals have very limited applications. Compared with nanofabricated porous materials with $p$ as tens of nanometers or longer, superlattice thin films can achieve sub-10 nm periodicity and atomically smooth interfaces, offering more opportunities in phononic studies. For 2D materials, the aspect ratio restriction for pore-drilling techniques is not applicable and sub-10 nm pitches can be potentially fabricated with EBL patterning. For suspended samples, both a FIB and electron beams can be directly used for the drilling. However, varied pore-edge defects can be induced and may diminish the coherent phonon transport.

Without pore drilling, direct growth of nanoporous films[39] or 2D materials with patterned pillar-like masks can minimize pore-edge defects to potentially yield a comparable case as superlattice thin films. For thermal management and thermoelectric applications, atomic-thick 2D materials are restricted by their low in-plane thermal and electrical conductances. Challenges also exist in the accurate thermal measurements of 2D structures.[140] In practice, ultrafine nanoporous thin films can be more useful. In this aspect, direct MOCVD growth may achieve a much higher aspect ratio for the through-film pores, compared with $t/d < 3$ for dry-etched films. The thickness of the grown film can be possibly larger than the height of the mask. Using only 40-nm-thick nanoporous $SiO_2$ as a mask, vertical micrometer-length GaN nanowires with a 50 nm diameter can be grown.[141] Considering the ~5 nm spatial resolution for the state-of-the-art EBL, sub-10 nm masks can thus be fabricated to directly grow ultrafine nanoporous thin films with a thickness of tens of nanometers. The minimum $k$ predicted for superlattices[24, 136] is anticipated for ultrafine porous films or 2D materials with smooth pore edges, resulting from the interplay of periodicity for coherent phonon transport and pore-edge defects for incoherent phonon transport.

Existing studies about phononic crystals are proposed for thermoelectric applications, in which bulk-like electrical properties can be maintained but the thermal conductivity $k$ can be dramatically reduced.[14, 31, 92] Other than thermoelectrics, heat guide and phonon focusing are also studied by patterning designed nanoporous regions across a Si film.[3] When impurity atoms are further introduced to scatter the short-wavelength phonons, it is possible to narrow down the phonon conduction to the middle-wavelength regime. This so-called thermocrystal[9, 16, 17] are proposed for heat waveguides, thermal lattices, heat imaging, thermo-optics, thermal diodes, and thermal cloaking.

Besides aligned or staggered circular pores, other porous patterns are also studied under the circumstance of incoherent phonon transport.[142, 143] When the porous patterns are designed to tune the phonon transmission by pore-edge phonon scattering, asymmetric phonon transmission along the forward and backward directions may lead to thermal rectification effects.[55, 144] The high performance of such a thermal rectifier requires strong ballistic transport within the nanoporous pattern and strong specular reflection by pore edges. The two requirements can be mostly satisfied at low temperatures, where dominant phonon MFPs become much longer than the structure size and the phonon wavelengths become long compared with the pore-edge roughness to ensure specular phonon reflection. Following this, Schmotz et al. reported thermal rectification effect at 150 K for designed porous patterns though this effect was not found at 300 K.[145] However, no thermal rectification effect beyond the instrumental uncertainty was found by Gluchko et al. at 4.2–300 K in Si films with Pacman pores[146] and by Kage et al. at 5–295 K in Si films with dog-leg-shaped pores.[147] In physics, Maznev et al.[148] argued that nonlinearity must be introduced into a system and asymmetric structure alone was not sufficient to induce thermal rectification. Different from a ballistic thermal rectifier, thermal rectification was reported by Wang et al. for suspended graphene with one end modified by nanostructured created with electron-beam-induced deposition or FIB drilling.[99] Thermal rectification was simply caused by the contrast in the temperature dependence of $k_L$ for the unpatterned and nanostructure-patterned regions. The total thermal conductance of the sample was thus different when the hot and cold ends were switched.[124]

In summary, coherent and incoherent phonon transport in periodic porous or further pillar-modified structures can play an important role in the manipulation of thermal transport within materials and devices. Despite numerous atomistic simulations, ultrafine periodic structures down to the atomic level are still hard to be fabricated and the amorphous pore edges often destroy coherent phonon phases. In experiments, coherent phonon transport and thus phononic effects can only be notable around or below 10 K, which largely restricts the applications of these materials. In this aspect, future work should be carried out on fabricating and measuring ultrafine nanoporous structures with minimized pore-edge defects. With such structures, existing designs of optical and acoustic devices can be adopted by phononic devices to manipulate phonons. Without phononic defects, classical phonon size effects are considered for such porous structures but the pattern shape and distribution can still be varied to benefit many applications.[55, 142] In general, certain

"thermally dead volume" can be introduced in a patterned thin film, where phonons can be trapped and contribute less to heat conduction. Examples can be found in nanoladders with a row of rectangular holes[143] and SiNWs with periodic wings.[149]

## Acknowledgements

Q.H. acknowledges the support from the U.S. Air Force Office of Scientific Research (award number FA9550-16-1-0025) for studies on nanoporous materials and National Science Foundation (grant number CBET-1651840) for phonon simulations. N.Y. was sponsored by National Natural Science Foundation of China (No. 51576076 and No. 51711540031), Natural Science Foundation of Hubei Province (2017CFA046) and Fundamental Research Funds for the Central Universities (2019kfyRCPY045). An allocation of computer time from the UA Research Computing High Performance Computing (HPC), High Throughput Computing (HTC) at the University of Arizona, the National Supercomputing Center in Tianjin (NSCC-TJ) and China Scientific Computing Grid (ScGrid) are gratefully acknowledged.